\documentclass[prl,twocolumn]{revtex4}
\usepackage{amssymb}
\usepackage{graphicx}
\usepackage{amsfonts}
\usepackage{amsmath}

\begin{document}

\title{Impurity effects on Fabry-P\'{e}rot physics of ballistic carbon nanotubes}

\author{ F. Romeo$^{1,2}$, R. Citro$^{1,2}$ and A. Di Bartolomeo$^{1,2}$}
\affiliation{$^{1}$Dipartimento di Fisica ''E. R. Caianiello'', Universit{\`a} degli Studi di Salerno and $^{2}$Institute CNR-SPIN,\\
  Via Ponte don Melillo, I-84084 Fisciano (Sa), Italy}


\begin{abstract}
We present a theoretical model accounting for the anomalous Fabry-P\'{e}rot pattern observed in the ballistic conductance of a single-wall carbon nanotubes. Using the scattering field theory, it is shown that the presence of a limited number of impurities along the nanotube can be identified by a measurement of the conductance and their position determined. Impurities can be made active or silent depending on the interaction with the substrate via the back-gate. The conceptual steps for designing a bio-molecules detector are briefly discussed.
\end{abstract}

\maketitle

\textit{Introduction -} Single-walled carbon-nanotubes (SWCNTs) are chiral graphene tubules able to show one-dimensional metallic or semiconducting properties depending on their diameter and helicity [\onlinecite{cnt}]. Such systems have unique electronic and mechanical properties allowing for the creation of devices useful for fundamental physics applications. Recently a variety of such nanotube-based devices, e.g. single-electron transistors [\onlinecite{set_cnt}], field-effect transistors [\onlinecite{fet_cnt}], memory cells [\onlinecite{dib}] and electromechanical devices [\onlinecite{mems_cnt}], have been realized. In the non-interacting resonant tunneling regime (i.e. for $U/\Gamma \ll 1$, $U$ and $\Gamma$ being the Coulomb interaction strength and the tunneling coupling to the leads, respectively), a SWCNT behaves like an almost one dimensional ballistic wire where the quantum interference and the Fabry-P\'{e}rot (FP) physics can be tested. \\
A FP interferometer based on individual SWCNT has been studied in Ref.[\onlinecite{fp_cnt}]. In such devices, the nanotube behaves as a coherent electron wave-guide where a resonant cavity is formed within nanotube-electrode interfaces. As a result of the electron scattering at the interfaces, the resonant cavity becomes populated by quasi-bound states formed by the propagating modes (left and right movers) inside the nanotube. The interference between left and right movers can be controlled through a back-gate $V_g$ which produces an electromagnetic phase shift responsible for the FP oscillations of the $G$-$V_g$ characteristic, $G$ being the differential conductance of the device. Even though the observation of the FP interference is reported in Ref.[\onlinecite{fp_cnt}] (see Fig. 1 of that work), the $G$ \textit{vs} $V_g$ curves show additional harmonics compared to the ones expected for an ideal FP interferometer. Recently, these anomalies have been interpreted as the signature of interaction effects [\onlinecite{interact_cnt}], while in the original work the magnitude and position of the conductance dips were attributed to disorder effects arising from the underlying substrate or due to intrinsic nanotube defects. The latter hypothesis has never been directly tested and thus a certain part of the experimental interpretation still needs clarification. \\
Moreover, a recent work[\onlinecite{bercioux}] showed the possibility to artificially induce defects in SWCNTs by $Ar^{+}$ irradiation. Such disordered 1D systems present random pattern of impurities able to confine the electrons which  exhibit \textit{particle-in-a-box} physics. \\
Motivated by the recent interest towards the few-defects physics in SWCNTs and in more controllable systems as e.g. cold atoms systems\cite{inguscio}, in this Letter we present a scattering field theory[\onlinecite{buttiker92}] of the FP physics in a one-dimensional wire in the presence of few defects. We would like to account for the  anomalies of the  $G$ \textit{vs} $V_g$ curves observed in Ref.[\onlinecite{fp_cnt}] following the original hypothesis of defect-induced modulation. Our analysis can be relevant to recognize and locate isolated defects in quasi-one-dimensional systems by a simple interferometry experiment.

\textit{Model and Formalism -} The low energy physics of a metallic impurity-free single-walled carbon nanotube is described by the Hamiltonian:
\begin{eqnarray}
H_0=-i\hbar v_F \sum_{\sigma,\alpha\in\{L,R\}}\int dx  \xi_{\alpha} : \psi^\dagger_{\alpha\sigma}(x)\partial_x \psi_{\alpha\sigma}(x):
\end{eqnarray}
where $\psi_{R(L)\sigma}$ is the right (left) mover electron field with spin $\sigma=\{\uparrow,\downarrow\}$ and chirality $\xi_R=1$ ( $\xi_L=-1$),
$x$ is the longitudinal coordinate, while $: \hat{\mathcal{O}}:$ stands for the normal ordering of the operator $\hat{\mathcal{O}}$ with respect to the equilibrium state defined by occupied energy levels below the Fermi sea. When the nanotube is put in contact with a metal to form the source/drain electrodes, the sites $x_{1/2}$ (located at nanotube/electrode interfaces) become scattering centers for the electrons and a backscattering occurs. Accordingly,  the following term is added to the Hamiltonian of the system\cite{note}:
\begin{eqnarray}
\label{eq:contacts}
H_{c}=\sum_{\sigma=\uparrow,\downarrow} \int dx &&\lbrack \Gamma_c(x) \psi^\dagger_{R\sigma}(x)\psi_{L\sigma}(x)+{\textit h.c.}
\rbrack,
\end{eqnarray}
where the function $\Gamma_c(x)=2 \hbar v_F \sum_{j=1,2}\delta(x-x_j)\lambda_j $ and $\lambda_j$ is the scattering strength. \\
The effect of a back gate on the electron transport can be taken into account by introducing a gate-dependent term:
\begin{eqnarray}
\label{eq:gates}
H_g=\sum_{\sigma}\int_{x_1}^{x_2} dx \lbrack e V_{g}(\rho_{R \sigma}+\rho_{L\sigma})\rbrack,
\end{eqnarray}
which depends on the electron densities $\rho_{\alpha \sigma}=:\psi^\dagger_{\alpha \sigma}\psi_{\alpha \sigma} :$  ($\alpha=L,R$ and spin $\sigma$) in the channel (i.e. the region $x_1< x <x_2$) and the electrostatic potential $V_g$
of the back-gate.\\
\begin{figure}
\centering
\includegraphics[clip,scale=0.55]{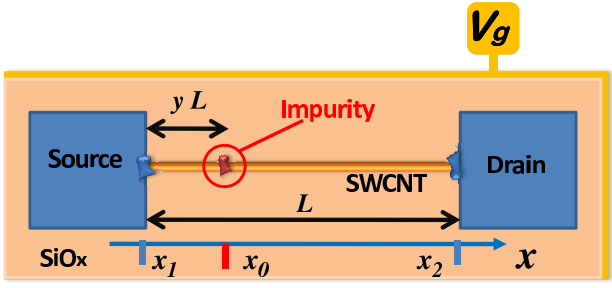}\\
\caption{Carbon nanotube based device described in the main text. The position $x_1$ and $x_2$ where the electrodes merge with the nanotube are scattering centers able to couple the left and right movers inside the channel. The back-gate $V_g$ controls the electromagnetic phase kept by the electrons interfering in the system, while the presence of an impurity in $x_0$ affects the Fabry-P\'{e}rot physics introducing additional harmonics to the G \textit{vs} $\Phi_g$ relation.}
\label{fig:device}
\end{figure}
The presence of an impurity with effective size $l_s$ comparable to the lattice constant $a_0$ along the nanotube affects the electron transport causing an additional scattering at $x_0 \in[x_1, x_2]$
which is described by the Hamiltonian:
\begin{eqnarray}
\label{eq:imp}
H_{I}=\sum_{\sigma=\uparrow,\downarrow} \int dx &&\lbrack \Gamma_I(x) \psi^\dagger_{R\sigma}(x)\psi_{L\sigma}(x)+ {\textit h.c.}
\rbrack,
\end{eqnarray}
where $\Gamma_I(x)=2 \hbar v_F \delta(x-x_0)\lambda_0 $.\\
To determine the differential conductance $G$ of the device a scattering field theory\textit{ \`{a }la } B\"{u}ttiker [\onlinecite{buttiker92}] can be implemented, while the space-time dependent electron fields can be obtained using the equation of motion (EOM) method.  In absence of impurities, i.e. when $\lambda_0 \rightarrow 0$, the dynamics of the electron field $\psi_{\alpha\sigma}(x)$ far from the contacts is described by the equation
\begin{eqnarray}
\label{eq:fields}
[\partial_t+\xi_{\alpha}v_F\partial_x-\frac{eV_g}{i\hbar}\Phi(x)]\psi_{\alpha\sigma}(x)=0,
\end{eqnarray}
where $\Phi(x)$ is a step-like function which takes value 1 in the channel, i.e. for $x_1<x<x_2$, and 0 elsewhere. The scattering fields solving (\ref{eq:fields})  are written below:
\begin{eqnarray}
\hat{\psi}_{R\sigma}(x,t)=\frac{e^{ik_E x-iEt/\hbar}}{\sqrt{2\pi\hbar v_F}}\times
                                                                             \begin{array}{cc}
                                                                               \hat{a}_{R\sigma}(E) & x<x_1\\
                                                                               e^{-i\frac{eV_g}{\hbar v_F}}\hat{c}_{R\sigma}(E) & x_1<x <x_2\\
                                                                               \hat{b}_{R\sigma}(E) & x>x_2 \\
                                                                             \end{array}
                                                                           \end{eqnarray}
and
\begin{eqnarray}
\hat{\psi}_{L\sigma}(x,t)=\frac{e^{-ik_E x-iEt/\hbar}}{\sqrt{2\pi\hbar v_F}}\times
                                                                             \begin{array}{cc}
                                                                               \hat{b}_{L\sigma}(E) & x<x_1\\
                                                                               e^{i\frac{eV_g}{\hbar v_F}}\hat{c}_{L\sigma}(E) & x_1<x <x_2\\
                                                                               \hat{a}_{L\sigma}(E) & x>x_2 \\
                                                                             \end{array}
                                                                          \end{eqnarray}
where $k_E=E/(\hbar v_F)$.
The interaction of the electrons with the delta-like potentials mimicking the contacts interface is described by imposing the appropriate boundary conditions of the scattering fields describing the incoming and the outgoing particles.
The $\mathcal{S}$-matrix relates the scattering fields according to $\hat{\textbf{b}}=\mathcal{S}\hat{\textbf{a}}$, where the outgoing (incoming) field vector is given by $\hat{\textbf{b}}=(\hat{b}_{L\uparrow}, \hat{b}_{L\downarrow}, \hat{b}_{R\uparrow}, \hat{b}_{R\downarrow})^t$ ($\hat{\textbf{a}}=(\hat{a}_{R\downarrow}, \hat{a}_{R\uparrow}, \hat{a}_{L\downarrow}, \hat{a}_{L\uparrow})^t$). The current operators within the scattering formalism can be written as follows:
\begin{eqnarray}
\hat{J}_L(t)&=&-(|e|/h)\sum_{i=1,2}[\hat{a}^{\dag}_i(t)\hat{a}_i(t)-\hat{b}^{\dag}_i(t)\hat{b}_i(t)]\\\nonumber
\hat{J}_R(t)&=&(|e|/h)\sum_{i=3,4}[\hat{a}^{\dag}_i(t)\hat{a}_i(t)-\hat{b}^{\dag}_i(t)\hat{b}_i(t)].
\end{eqnarray}
Considering that $\hat{a}_i(t)=\int dE e^{-iEt/\hbar}\hat{a}_i(E)$ and using the quantum average $\langle \hat{a}^{\dag}_j(E)\hat{a}_i(E')\rangle=\delta_{ij}\delta(E-E')f_i(E)$ ($f$ being the Fermi function) for the incoming fields, one obtains the linear response charge current $J_L$ measured in the left lead:
\begin{equation}
\label{eq:left_current}
J_L=-(e^2/h)\int d\xi \sum_{i,j}\Bigl[\delta_{ij}-|\mathcal{S}_{ij}(\xi)|^2\Bigl](\partial_{\xi} f(\xi))_{eq}V_j,
\end{equation}
where the voltage bias $V_j=V$ ($-V$) if $j=1,2$ $(j=3,4)$, $i\in\{1, 2\}$, $j\in\{1,...,4\}$, while the indices $(i,j)$ are defined in \cite{note2}.
Starting from Eq.(\ref{eq:left_current}) the differential conductance $G=\partial_V J_L$ is easily obtained. In the case of a single delta-like scatterer in the wire (mimicking the contact interface or an impurity) the only nonvanishing elements of the scattering matrix are: $\mathcal{S}_{12}=\mathcal{S}_{21}=\mathcal{S}_{34}=\mathcal{S}_{43}=-2i\gamma/(1+\gamma^2)$, $\mathcal{S}_{14}=\mathcal{S}_{41}=\mathcal{S}_{23}=\mathcal{S}_{32}=(1-\gamma^2)/(1+\gamma^2)$, where $\gamma=\lambda_j$. The formalism described so far can be easily generalized to the \textit{n}-impurities case by composition law of scattering matrices as given e.g. in Ref.[\onlinecite{ouisse_book}]. In this way the charge transport  through a SWCNT can be studied in the quasi-ballistic regime. Below we focus on the case of three scatterers in the zero-temperature limit.\\
\textit{Results -} Our approach is employed to analyze the experimental data presented in Ref.[\onlinecite{fp_cnt}] where the FP physics has been shown by the measure of the differential conductance of a single-walled carbon nanotube.
For an open ballistic channel the zero-bias conductance is $N G_0$ where $G_0=e^2/h$ is the quantum of conductance and $N$ the number of modes. Thus one expects that in the absence of scattering events the ballistic conductance of a SWCNT is $4e^2/h$; however in the presence of non-ideal contacts or of the interaction with substrate impurities, back-scattering events reduce the transmission probability of the system. In order to understand such effects on the differential conductance we take $\lambda_0=\lambda_1=\lambda_2=\gamma$ and focus on the high transparency limit (i.e. $\gamma\ll 1$) which seems to be appropriate for the experiment reported in Ref.[\onlinecite{fp_cnt}]. Under these assumptions the analytical expression of the conductance to the order $\gamma^2$ is
\begin{eqnarray}
\label{eq:conductance_approx}
G/G_0 & \simeq & 4-16 \gamma^2\Bigl[3+2\cos(2\Phi_g)+\\\nonumber
&+& 2\cos(2\Phi_g (\textsl{y}-1))+ 2\cos(2 \Phi_g \textsl{y} )\Bigl]+\mathcal{O}(\gamma^4),
\end{eqnarray}
where the gate-induced electromagnetic phase is defined as $\Phi_g=eV_g L/(\hbar v_F)$, while $\textsl{y}$ is a dimensionless factor measuring the position  of the impurity normalized to the total length $L$ of the nanotube. Eq.(\ref{eq:conductance_approx}) shows that the FP physics is controlled by the back-gate $V_g$.
 Differently from the impurity-free case, the gate-induced oscillations of $G(\Phi_g)$ include the
  terms $\cos(2\Phi_g \beta)$, $\beta \in \{\textsl{y},\textsl{y}-1\}$, which collapse into the usual $\cos(2\Phi_g)$ when $\beta=0$ or $1$, i.e. for an impurity residing on the left or the right contact. The properties reported above imply  that a carbon-nanotube with an impurity behaves like a sequence of two FP interferometers of length $\textsl{y} L$ and $(1-\textsl{y})L$, respectively. Since a sequence of \textit{n}  FP interferometers can be used to design systems with special optical properties, one can expect that controlling the introduction of impurities in ballistic nanotubes can be useful in obtaining quantum devices with controllable electronic properties.
  A work on the effects of defects produced in CNT by means of $Ar^{+}$ irradiation recently appeared[\onlinecite{bercioux}],
  even though only random patterns of impurities were obtained. \\
In Fig.\ref{fig:fig2ab} we compare our results with the FP pattern observed in Fig.1 of Ref.[\onlinecite{fp_cnt}] and reproduced in the lower panel of Fig.\ref{fig:fig2ab}. In order to achieve a quantitative comparison with experimental data, we introduce the contacts conductance, i.e. $G_1$ and $G_2$, which takes into account the renormalization of the electrons tunneling in the nanotube, while the conductance $G$ of the whole system is obtained by the composition law:
\begin{equation}
\label{eq:conductance_renormal}
G=\frac{G_{NT}}{1+z G_{NT}},
\end{equation}
where $G_{NT}$ is the differential conductance (in units of $G_0=e^2/h$) of the SWCNT as computed using the previous scattering model, while $z$ is the
contacts transparency, $z=G_0(G_1+G_2)(G_1 G_2)^{-1}$. In the upper panel of Fig.\ref{fig:fig2ab} we report the differential conductance $G$ (in unit of $e^2/h$) \textit{vs} $V_g$ using the following set of parameters: $\gamma=0.05$, $y=0.4465$, $\Phi_g=\Omega_V(V_g-V^{0}_g)$, $\Omega_V=4.65$$/V$, $V^{0}_g=-2.07$ $V$, $z=0.057$. By comparing the upper and lower panel of Fig.\ref{fig:fig2ab} the modulation of the maxima and minima, the oscillation period and the general qualitative features of the experimental
$G$ \textit{vs} $V_g$ curve are correctly reproduced.
\begin{figure}[h]
\centering
\includegraphics[clip,scale=0.55]{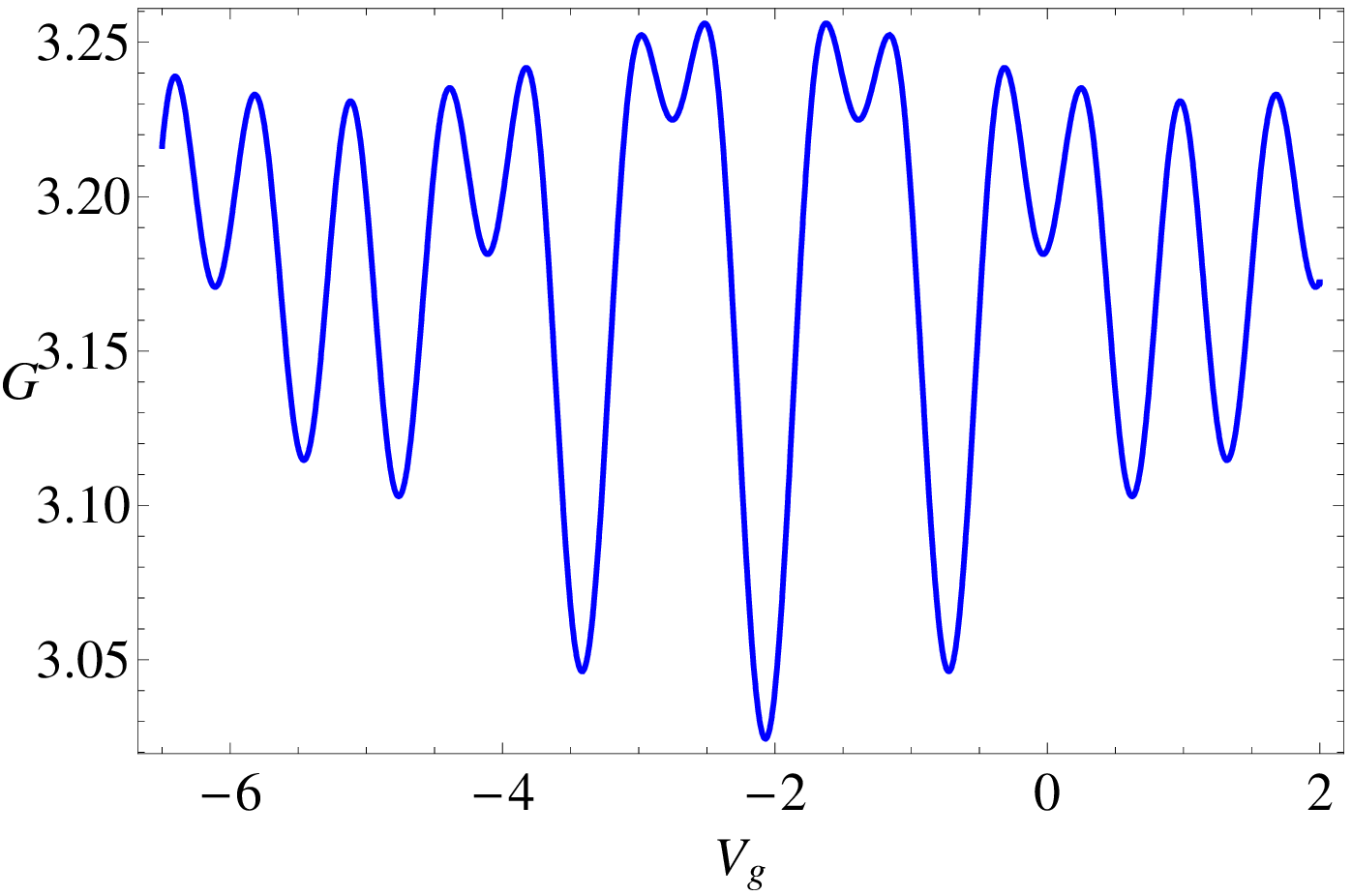}\\
\includegraphics[clip,scale=0.55]{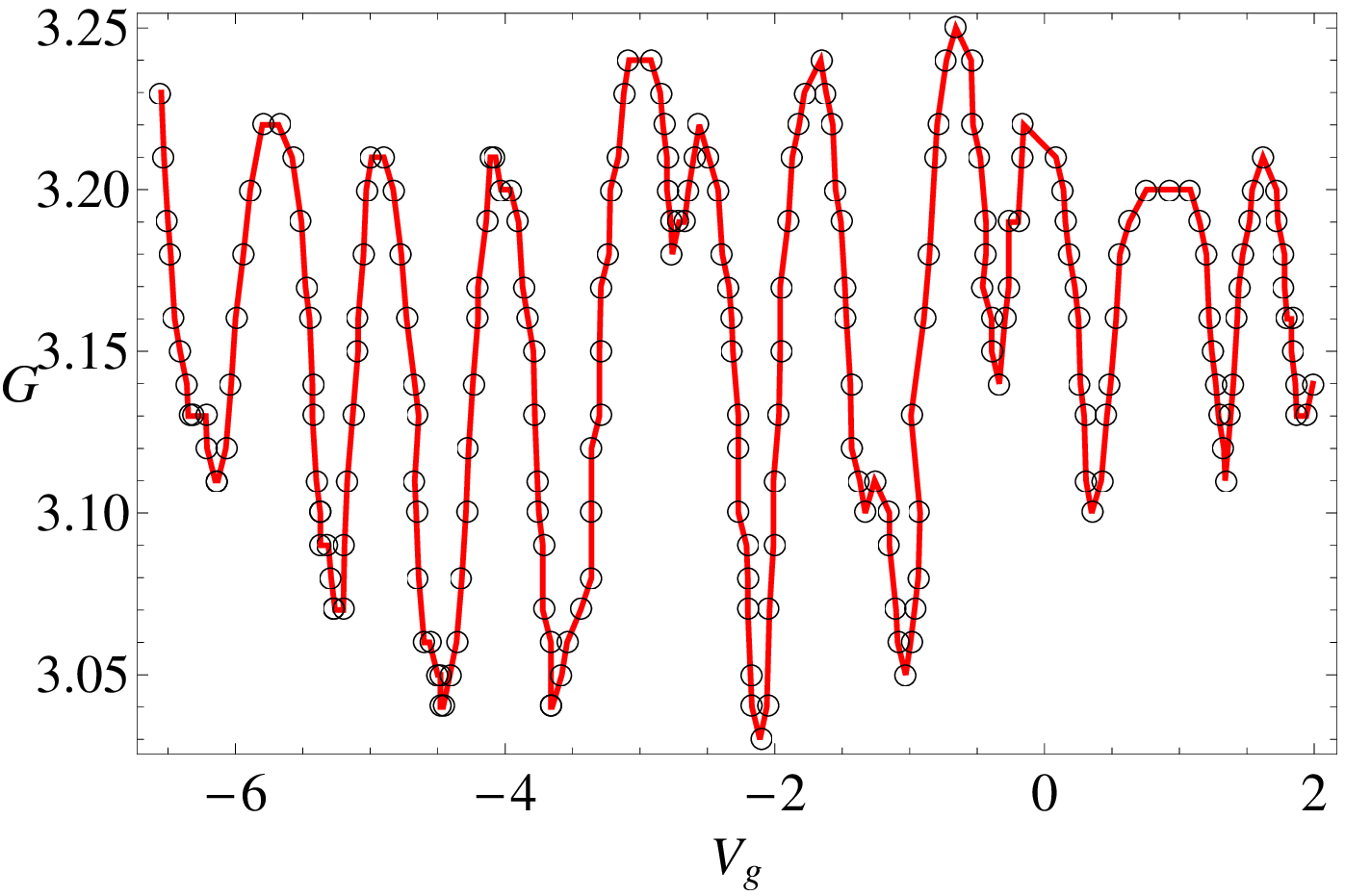}\\
\caption{Upper panel: Differential conductance $G$ (in unit of $e^2/h$) \textit{vs} $V_g$ using the following parameters: $\gamma=0.05$, $y=0.4465$, $\Phi_g=\Omega_V(V_g-V^{0}_g)$, $\Omega_V=4.65$$/V$, $V^{0}_g=-2.07$ $V$, $z=0.057$. Lower panel: Experimental data as extracted by Fig.1 of Ref.[\onlinecite{fp_cnt}].}
\label{fig:fig2ab}
\end{figure}
As for the fitting parameters, $z=0.057$ corresponds to a contact conductance $G_1\simeq G_2=G_c \approx 30 e^2/h$. This value is consistent with the one usually found in carbon-based mesoscopic systems [\onlinecite{contact_res_cnt}]. The other fitting parameter is $\textsl{y}$ which fixes the position of the scattering center at distance $0.447$$L$ from the left lead (see Fig.\ref{fig:device}), while the scattering strength is $\gamma=0.05$, which is quite weak and consistent with the high value of the differential conductance of the device. Let us note that in general the scattering strength of the impurities depends on the back-gate $V_g$ [\onlinecite{v_imp_cnt}], and thus by varying $V_g$ one can observe the activation (switching off) of an impurity previously silent (active). This phenomenon can be detected by the deformation of the interference pattern and the appearance of additional unusual features.
In the above analysis the relation between the electromagnetic phase $\Phi_g$ and the back-gate $V_g$ has been taken of the form $\Phi_g=\Omega_V(V_g-V^{0}_g)$ to describe the linear electromagnetic response of the substrate to the applied potential $V_g$ around the
working point defined  by $V^{0}_g$.
Indeed, $V_g-V^{0}_g$ represents an effective potential that takes into account charge trapping at the oxide-nanotube interface.
When the parameter $\Omega_V$ is compared to $eL/(\hbar v_F)$, i.e. the value of $\Phi_g/V_g$,
the relation $\Omega_V=eL\eta/(\hbar v_F)$ is correctly found. The dimensionless factor  $\eta \approx 10^{-2}$ is known as gate efficiency[\onlinecite{cnt_book}] and represents the fraction of applied back-gate voltage felt by the SWCNT electrons.
\begin{figure}
\centering
\includegraphics[clip,scale=0.55]{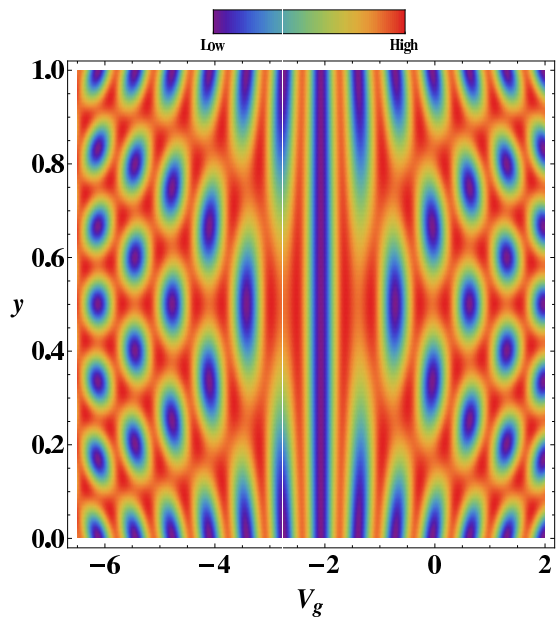}\\
\caption{Density plot of the differential conductance $G$ (in unit of $e^2/h$) in the plane $(V_g, y)$ obtained by fixing the model parameters as done in Fig.\ref{fig:fig2ab}. One can notice the modulation of the FP pattern as a function of the impurity position $\textsl{y}$.}
\label{fig:fig3}
\end{figure}
In Fig.\ref{fig:fig3} we report the density plot of the differential conductance $G$ (in unit of $e^2/h$) in the plane $(V_g, y)$ obtained by fixing the model parameters as done in Fig.\ref{fig:fig2ab} and varying the impurity position. As evident from the analysis of that figure, the FP pattern is strongly affected by the impurity position $\textsl{y}$.  The impurities located close to the leads, i.e. for $0<y<0.05$ or $0.95<y<1$, only cause a lowering of the effective contacts transparency without introducing important changes in the FP curves. Impurities located in the channel, i.e. for $0.2<y<0.8$, induce modifications in the $G$ \textsl{ vs} $V_g$ curves (as the location and amplitude of the dips) and thus their presence can be easily detected. The mentioned properties could also be useful to design electrically-readable sensors where a certain number of sites along the nanotube behave like \textit{instructed impurities}. Such impurities can be of two types, active or silent, depending on the interaction with the environment (chemical or biological)[\onlinecite{chem_sensor_cnt}].
Finally we point out that the results shown in Fig.\ref{fig:fig3} are similar to the ones obtained in Ref.[\onlinecite{bercioux}] where the atomic structure and chirality of the SWCNT were fully taken into account.\\
\textit{Conclusions -} We presented a scattering theory \textit{\`{a} la} B\"{u}ttiker accounting for the FP pattern measured in ballistic SWCNTs with a limited number of impurities. Using our theoretical model,
we were able to reproduce the experimental data reported in Ref.[\onlinecite{fp_cnt}] offering a quantitative test for an
impurity-modulated FP pattern in the $G$-$V_g$ characteristic. We provided an alternative explanation of the experimental data
compared to the ones given in Ref.[\onlinecite{interact_cnt}]. Thus, the present work adds
supplementary information useful in understanding interferences effects in carbon-based resonant cavities. On the pratical point of view,
the sensible dependence of the differential conductance on the impurities position could be a useful ingredient in designing biological sensors.\\
\textit{Acknowledgements -} The authors wish to acknowledge D. Bercioux for helpful discussions and Prof. H. Park for having given access to the
experimental data.
\bibliographystyle{prsty}

\begin{thebibliography}{99}
\bibitem{cnt}S. Iijima, Nature (London) \textbf{354}, 56 (1991); Noriaki Hamada, Shin-ichi Sawada, and Atsushi Oshiyama, Phys. Rev. Lett. \textbf{68}, 1579 (1992); C. Sch\"{o}nenberger, Semicond. Sci. Technol. \textbf{21}, S1-S9 (2006)
\bibitem{set_cnt}M. Bockrath \textit{et al.}, Science \textbf{275}, 1922 (1997); S. J. Tans \textit{et al.}, Nature \textbf{386}, 474 (1997).
\bibitem{fet_cnt}S. J. Tans, R. M. Verschueren, C. Dekker, Nature \textbf{393}, 49 (1998); R. Martel \textit{et al.}, Appl. Phys. Lett. \textbf{73}, 2447 (1998); A. Di Bartolomeo, M. Rinzan, A. K. Boyd, Y. Yang, L. Guadagno, F. Giubileo, and P. Barbara, Nanotechnology \textbf{21}, 115204 (2010)
\bibitem{dib}A. Di Bartolomeo, Y. Yang, M. B. M. Rinzan, A. K. Boyd, P. Barbara, Nanoscale Res. Lett. \textbf{5}, 1852 (2010)
\bibitem{mems_cnt}T. W. Tombler \textit{et al.}, Nature \textbf{405}, 769 (2000); T. Rueckes \textit{et al.}, Science \textbf{289}, 94 (2000)
\bibitem{fp_cnt}W. Liang, M. Bockrath, D. Bozovic, J. H. Hafner, M. Tinkham, and H. Park, Nature \textbf{411}, 665 (2001)
\bibitem{interact_cnt}Claudia S. Pe\c{c}a, Leon Balents, and Kay J\"{o}rg Wiese, Phys. Rev. B \textbf{68}, 205423 (2003)
\bibitem{bercioux}G. Buchs, D. Bercioux, P. Ruffieux, P. Gr\"{o}ning, H. Grabert, and O. Gr\"{o}ning, Phys. Rev. Lett. \textbf{102}, 245505 (2009);
D. Bercioux, G. Buchs, H. Grabert, and O. Gr\"{o}ning, Phys. Rev. B \textbf{83}, 165439 (2011)
\bibitem{inguscio} J. Catani \textit{at al.}, \texttt{arXiv: 1106.0828v1} (2011)
\bibitem{buttiker92}M. B\"{u}ttiker, Phys. Rev. B \textbf{46}, 12485 (1992)
\bibitem{note} Note that we neglect the renormalization of particles incoming from the electrode due to the metal-nanotube interaction, that will be taken into account in the calculation of $G$ afterwards (see Eq. (\ref{eq:conductance_renormal})).
\bibitem{note2}Notice that for the outgoing states the indices of the scattering matrix have to be interpreted as follows: $i=1 \rightarrow (L,\uparrow)$, $i=2 \rightarrow (L,\downarrow)$, $i=3 \rightarrow (R,\uparrow)$,
$i=4 \rightarrow (R,\downarrow)$. Concerning the incoming states the notation becomes: $i=1 \rightarrow (R,\downarrow)$, $i=2 \rightarrow (R,\uparrow)$, $i=3 \rightarrow (L,\downarrow)$, $i=4 \rightarrow (L,\uparrow)$.
\bibitem{ouisse_book} T. Ouisse, \textit{Electron Transport in Nanostructures and Mesoscopic Devices: An Introduction}, (John Wiley and Sons Inc, Hoboken, NJ, 2008); see chap. 4.
\bibitem{contact_res_cnt}Seong Chu Lim \textit{et al.}, Appl. Phys. Lett. \textbf{95}, 264103 (2009)
\bibitem{v_imp_cnt}M. Bockrath \textit{et al.}, Science \textbf{291}, 283 (2001)
\bibitem{cnt_book}F. L\'{e}onard, \textit{The Physics of Carbon Nanotube Devices}, (William Andrew Publishing , Norwich, NY, 2009)
\bibitem{chem_sensor_cnt}See for instance Jing Kong \textit{et al.}, Science \textbf{287}, 622 (2000)
\end{thebibliography}

\end{document}